\documentclass[journal, 10pt]{IEEEtran}
\usepackage[utf8]{inputenc}
\usepackage{graphicx}
\usepackage{xcolor}
\usepackage{multirow}
\usepackage{amsmath,amsfonts,amssymb,amstext}
\usepackage{float}
\floatstyle{plaintop}
\restylefloat{table}
\usepackage{hyperref}
\usepackage{multirow}

\title{Despeckling Polarimetric SAR Data Using a Multi-Stream Complex-Valued  Fully Convolutional Network  \thanks{\noindent Adugna G.~Mullissa and Johannes Reiche are with the Laboratory of Geo-information Science and Remote sensing, Wageningen University, 6700 AA  Wageningen, The Netherlands. Claudio Persello is with the Faculty of Geo-information Science and Earth Observation (ITC) of the University of Twente.  (email: adugna.mullissa@wur.nl,c.persello@utwente.nl and johannes.reiche@wur.nl). }}
\author{Adugna G. Mullissa,~\IEEEmembership{Member, IEEE}, Claudio Persello,~\IEEEmembership{Senior Member, IEEE} and Johannes Reiche }

\newcommand{\adugna}[2]{\textcolor{blue}{#2}}

\begin{document}

\maketitle

\begin{abstract}

\adugna{}{Note: this is a pre-print version of our work accepted for publication in IEEE Geoscience and Remote Sensing Letters. } A Polarimetric Synthetic Aperture Radar (PolSAR) sensor is able to collect images in different polarization states, making it a {rich source of information} for target characterization. PolSAR images are inherently affected by speckle. Therefore, before deriving ad-hoc products from the data, the polarimetric covariance matrix needs to be estimated by reducing speckle. In recent years, deep learning based despeckling methods have started to evolve from single channel SAR images to PolSAR images. To this aim, deep learning based approaches separate the real and imaginary components of the complex-valued covariance matrix and use them as independent channels in a standard convolutional neural networks. However, this approach neglects the mathematical relationship that exists between the real and imaginary components, resulting in sub-optimal output. Here, we propose a multi-stream complex-valued fully convolutional network  (CV-deSpeckNet\footnote{\url{https://github.com/adugnag/CV-deSpeckNet}}) to reduce speckle and effectively estimate the  PolSAR covariance matrix. To evaluate the performance of CV-deSpeckNet, we used Sentinel-1 dual polarimetric SAR images to compare against its real-valued counterpart, that separates the real and imaginary parts of the complex covariance matrix. {CV-deSpeckNet was also compared against the state of the art PolSAR despeckling methods}. The results show CV-deSpeckNet was able to be trained with a fewer number of samples, has a higher generalization capability and resulted in a higher accuracy than its real-valued counterpart {and state-of-the-art PolSAR despeckling methods}. {These results showcase the potential of complex-valued deep learning for PolSAR despeckling.}
\end{abstract}


\IEEEpeerreviewmaketitle

\section{Introduction}

The advent of freely available multiple polarization SAR images, such as Sentinel-1 dual polarimetric SAR (PolSAR) images, has been a game changer for all-weather day/night geospatial applications.  However, the exploitation of these data sets is complicated by the presence of speckle. Speckle is the effect that occurs from the interference of backscattered signals from multiple individual scatterers within a resolution cell. In polarimetric SAR, due to the presence of speckle, the main interest is not in the scattering matrix itself, but the estimated covariance matrix that determines the randomness  of  the  acquired  SAR  data  vector.  The covariance matrix defines the polarimetric properties of the image and  has to be estimated first to derive ad-hoc products such as target decomposition and terrain classification \cite{lee1999polarimetric}.\\

\noindent  Most polarimetric covariance matrix estimation methods proposed in the literature focus on spatially adaptive filters defined in a neighborhood window \cite{lee1999polarimetric} \cite{lee2009polarimetric}. Therefore, the main challenge is selecting which pixels to average together and how to assign the weight to each pixel. Blind low pass filters such as the boxcar filters  are ineffective in preserving resolution, edges and point scatterers in the PolSAR data.  Lee et al. \cite{lee1999polarimetric} improved these drawbacks by minimizing the mean square error of the trace of the covariance matrix in a series of edge aligned windows to filter elements of the covariance matrix. In \cite{lee2005scattering} and \cite{mullissa2017scattering} the scattering mechanisms is determined on a pixel to pixel basis to establish similarity. In \cite{lee2017polarimetric}, the authors addressed the bias issue observed with previous filters by redefining the range based on speckle probability function. Deladalle et al. \cite{deledalle2014nl} used a non-local means approach to accurately estimate the covariance matrix in a heterogeneous medium without losing resolution. Recently, Deladalle et al.  \cite{deledalle2017mulog} used a homomorphic approach to convert the PolSAR signal to an additive noise model that embeds a Gaussian denoiser to effectively filter speckle and estimate PolSAR covariance matrix in a heterogeneous medium. \\

\noindent Recently, deep learning based single polarization SAR image despeckling techniques have gained attention \cite{chierchia2017sar} \cite{zhang2018learning}. These methods operate by feeding pairs of noisy and clean images in the deep learning network so that the network learns a non-linear function to transform the noisy input images to the filtered output. Deep learning based polarimetric covariance matrix estimation is understudied compared to the single channel SAR despeckling and only a few studies applied convolutional neural network (CNN) to despeckle PolSAR data. Pan et al. \cite{pan2019filter} used a pretrained Gaussian denoising network to filter fully polarimetric SAR data and reported promising results. However, these methods did not consider the complex-valued nature of the covariance matrix. They separated the real and imaginary parts of the off diagonal elements of the covariance matrix as real channels, thereby neglecting the mathematical relationship that existed between them.   \\ 

\noindent To overcome these limitations and learn a more robust feature representation, complex-valued neural networks offer a potential solution.  Initial works on complex-valued neural networks focused on leveraging the properties of complex numbers to learn a more robust transformation functions  than real-valued networks \cite{nitta1997extension}  \cite{hirose2013complex}. However, these networks provided the theoretical basis for complex-valued neural network{ and did not see much real world applications}. Recent complex-valued deep learning frameworks focused on replicating the success of real-valued CNN. Therefore, these networks maintained the standard deep neural network architecture but redesigned the building blocks to accommodate complex-valued data tensors \cite{trabelsi1705deep}. In the PolSAR domain, these type of complex-valued CNN's were applied to accurately classify polarimetric SAR images \cite{zhang2017complex} \cite{mullissa2019polsarnet}. To the best knowledge of the authors, the application of a complex-valued deep neural network for despeckling PolSAR data is yet to be demonstrated. In this letter,  we propose for the first time a new architecture named CV-deSpeckNet that is designed  to estimate a dual polarimetric covariance matrix in the complex domain.

\section{Method}

\subsection{SAR Polarimetry}
\noindent A data vector in fully polarimetric SAR sensors assuming reciprocity is given as: $k = \frac{1}{\sqrt{2}}\left[ \begin {array}{ccc}
S_{{YY}}+S_{{XX}}
     & S_{{YY}}-S_{{XX}} & 2S_{{XY}}
\end{array} \right]^T,$ where  $^T$ designates a vector transpose. In dual polarimetric data as in Sentinel-1 configuration (VV VH), $k$ reduces to: ${k = \left[ \begin {array}{ccc}
S_{{XX}} & 2S_{{XY}} \end{array} \right]^T}.$  Where the complex scattering coefficient  $S_{XY}$ indexed as $X,Y = (V,H)$  represents the vertical ($V$) and horizontal ($H$) polarization states \cite{mullissa2018polarimetry}. In a distributed medium, $k$ follows a zero mean  multivariate complex circular Gaussian probability density function (pdf) given as,

\begin{equation}
\displaystyle
p(k) = \frac{1}{\pi^3 |C|} \mathrm{exp}(-k ^\dagger C^{-1} k),
\end{equation}

\noindent is insufficient to describe the scattering process of the scene. Therefore, the second order statistics represented by the covariance matrix $C$ that define the pdf of $k$ is estimated to describe the scattering process. Here, $C$ is given as: $C = E \lbrace k k^\dagger \rbrace$, where $E \lbrace \rbrace$ is the expectation operator, $|C|$ is the determinant of $C$ and $^\dagger$ is the matrix conjugate transpose.  Here, $C$ is an unknown deterministic quantity that has to be estimated from the data and follows a complex Wishart distribution \cite{lee2017polarimetric}. In the PolSAR speckle filtering literature the expectation operator is replaced by spatially adaptive filters assuming stationarity and ergodicity within the selected pixels. \\

\noindent The observed covariance matrix $\hat{C}$ follows a multiplicative noise model, i.e. $\hat{C}= C^{1/2} N C^{1/2}$, where $N$ is random speckle. The signal dependent multiplicative noise can be converted to a signal independent additive noise by taking the matrix logarithm of $\hat{C}$. The signal independent model can be converted back to its original multiplicative form by taking the matrix exponent.  In this letter, we apply a complex-valued deep fully convolutional network to estimate the covariance matrix  $\tilde{C}$.

\subsection{Network Architecture}

\begin{figure}[t!]
\centerline{\includegraphics[width=0.50\textwidth]{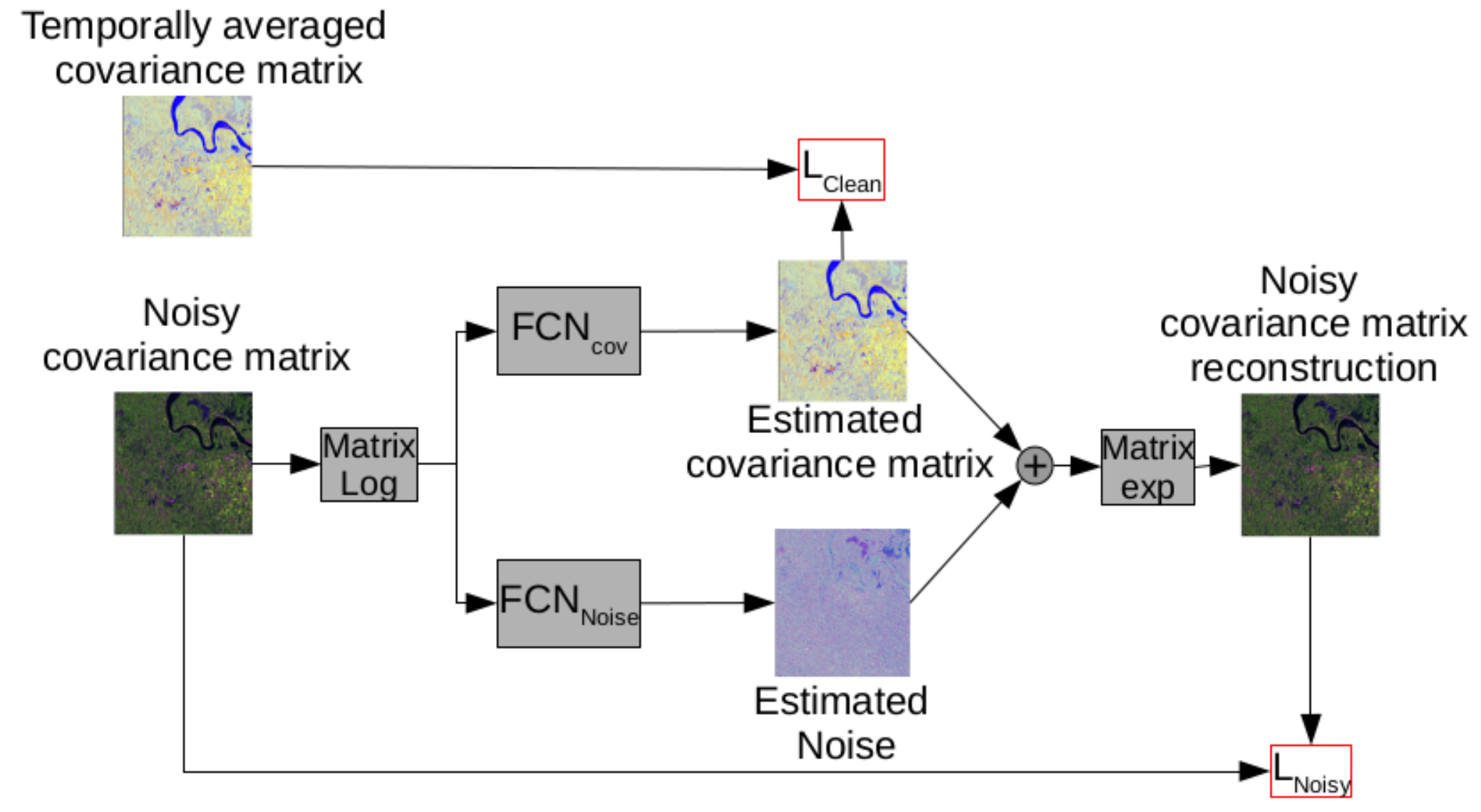}}
\caption{The architecture of CV-deSpeckNet. {The polarimetric covariance matrix in log form follows an additive noise model. Therefore the architecture of CV-deSpeckNet follows this model using two separate FCNs to estimate 1) the clean covariance matrix, and 2) the noise, separately. These two components are then summed up and converted to the multiplicative noise model by taking a matrix exponent to reconstruct the original noisy covariance matrix. At test time, only the clean covariance matrix estimation part ($\mathrm{FCN_{cov}}$) of the network is utilized.} }
 \label{fig:Fig2}
 \end{figure}

\noindent {We propose a multi-stream complex-valued  architecture named CV-deSpeckNet to estimate the covariance matrix $\tilde{C}$ and the noise $\tilde{N}$, separately (Figure~\ref{fig:Fig2}). A variant for single channel real-valued SAR intensity image despeckling was proposed in \cite{mullissa2020despecknet}.}  We adopted this architecture to train a robust model for learning feature representation and the underlying noise distribution in a complex-valued data, i.e. two identical fully convolutional networks (FCN) estimate $\tilde{C}$ ($\mathrm{FCN_{cov}}$) and $\tilde{N}$ ($\mathrm{FCN_{noise}}$), separately. With those two components, we reconstruct the noisy covariance matrix ($\hat{C}$) using the assumed signal independent additive noise that results from taking the matrix logarithm of $\hat{C}$. {CV-deSpeckNet is designed to rectify the limitation of deSpeckNet \cite{mullissa2020despecknet} by processing naturally complex-valued data in its native form without splitting real and imaginary components as separate channels.}\\

\noindent  Both ($\mathrm{FCN_{cov}}$) and  ($\mathrm{FCN_{noise}}$) consist of three main building blocks: 1) complex-convolution (CV-Conv) \cite{trabelsi1705deep}, 2) complex-activation (CReLU) \cite{mullissa2019polsarnet} and 3) complex-batch normalization (CV-BN) \cite{trabelsi1705deep}. The architecture {does not} use any pooling layers to avoid upsampling layers to reconstruct the feature maps to their original sizes, as these lay additional computational burden. Instead, we maintained the sizes of feature maps in the intermediate layers and increased the depth of the network. \\

\noindent  The deep learning objective is therefore formulated as:

\begin{equation}
\displaystyle
 \mathrm{arg}\min_{\theta}  \sum L(f_{\theta}(\hat{C}), C).
\end{equation}

\noindent Here, $\hat{C}$ is the observed rank 1 covariance matrix,  ${C}$ is the reference covariance matrix used as a proxy for the noise-free covariance matrix and $f_\theta$ is the deep neural network parametrized by the complex-valued parameters $\theta$ learned under the loss function $L$.\\

\noindent To train the network, we employ two types of loss functions, the sum squared error (SSE)-based $L_{cov}$ and $L_{noise}$ which are combined as, $L = L_{cov} + L_{noise}$.  \\

\noindent We apply the $L_{cov}$ loss between the reconstructed clean covariance matrix ($\tilde{C}$) and the reference temporally averaged  covariance matrix (${C}$) in the log domain.\\

\noindent The SSE  loss function in the complex domain is identical to its real-valued counterpart given as:
\begin{equation}
L_{cov}(\tilde{C},{C}) =  {\mu} \sum_{i=1}^H \sum_{j=1}^W \sum_{d=1}^D \frac{1}{2}({\tilde{C}_{ijd}} - {C}_{ijd})^2,
\end{equation}

\noindent where $H$, $W$ and $D$ are the height, width and number of feature maps in the estimated covariance matrix respectively, and $\mu$ is the weight assigned to the loss. Once $\tilde{C}$ and $\tilde{N}$ are reconstructed, an element-wise addition is applied and the reconstructed noisy covariance matrix is converted to the original linear scale by taking the matrix exponent (Figure~\ref{fig:Fig2}). The reconstructed noisy covariance matrix is then compared to the input covariance matrix by using another SSE-based loss $L_{noise}$:

\begin{equation}
L_{noise}(\hat{\tilde{C}},\hat{C}) =  {\xi} \sum_{i=1}^H \sum_{j=1}^W \sum_{d=1}^D \frac{1}{2}({\hat{\tilde{C}}_{ijd} - \hat{C}_{ijd})^2},
\end{equation}

\noindent Here, $\xi$ is the weight assigned to $L_{noise}$.

\section{Datasets}

\begin{figure}
	\centering
	
		\begin{tabular}{cc}
		\includegraphics[width=0.2\textwidth]{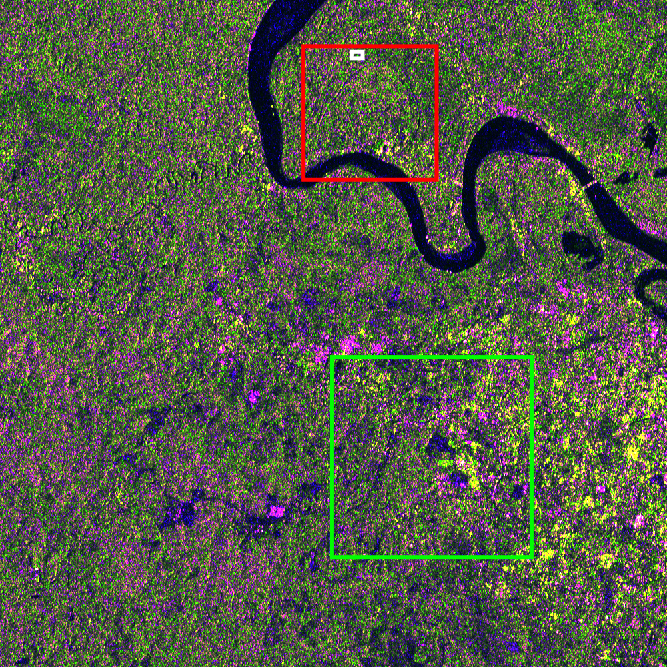}&
		\includegraphics[width=0.2\textwidth]{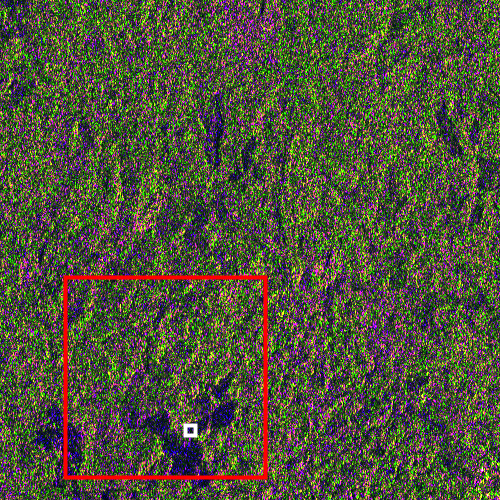}\\

	{(a)}&{(b)} 
		
	\end{tabular}

	\caption{{Input images used to test CV-deSpeckNet. The areas in the red and green boxes were used for qualitative comparison of the methods. The areas in the small white boxes in both images were used to estimate ENL (red: $C_{11}$, green: $C_{22}$, blue: $C_{11}/C_{22}$) (a) Test image 1, (b) Test image 2 }}\label{fig:Fig5}
\end{figure}

\noindent CV-deSpeckNet is tested using a rank 1 covariance matrix synthesized by taking the outer product of $k$ from a dual polarization Sentinel-1 single look complex (SLC) images acquired near the city of Jambi and the village of Tempino-ketjil, Sumatra, Indonesia. The images are acquired in C band in the interferometric wide swath mode (IW) in descending orbit. The images were acquired with an incidence angle of $40.02^0$ and have a resolution of $3.14\mathrm{m}\times 11.05\mathrm{m}$.  The image acquired around Jambi on May 13, 2019 and June 30, 2019 hereafter referred to as training image was used as a training data and each spanned $1,000 \times 1,000$ pixels. The trained model was tested on the image acquired on October 4, 2019 in the same geographical area as the training image hereafter referred to as test image1. In addition, an image acquired around Tempino-ketjil on May 13, 2019 hereafter referred to as test image 2 was used as a second test image. This image spanned $500 \times 500$ pixels. For synthesizing the reference labels we used 18 images acquired from May 13, 2019 to November 21, 2019 over the same area. {We used test image 1 and 2 for testing and obtaining the reported quality measurements}. Test image 1 covers a mixed urban and natural scene whereas the test image 2 covers a natural environment  (Figure~\ref{fig:Fig5}).  

\begin{figure*}
	\centering
	\begin{tabular}{cccccc}
		\includegraphics[width=0.15\textwidth]{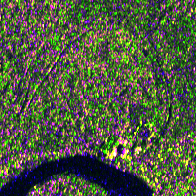}&
		\includegraphics[width=0.15\textwidth]{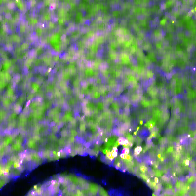}&
		\includegraphics[width=0.15\textwidth]{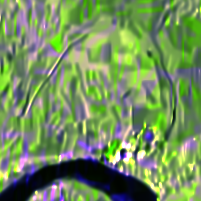}&
		\includegraphics[width=0.15\textwidth]{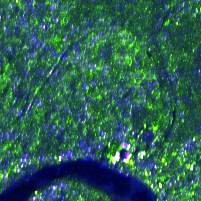}&
		\includegraphics[width=0.15\textwidth]{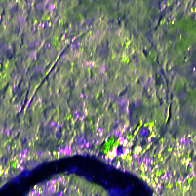}&
		\includegraphics[width=0.15\textwidth]{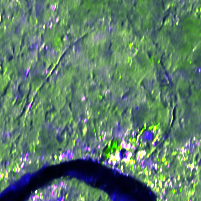}\\
		
	\end{tabular}

		\begin{tabular}{cccccc}
		\includegraphics[width=0.15\textwidth]{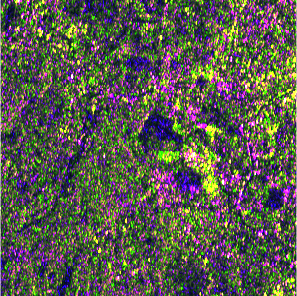}&
		\includegraphics[width=0.15\textwidth]{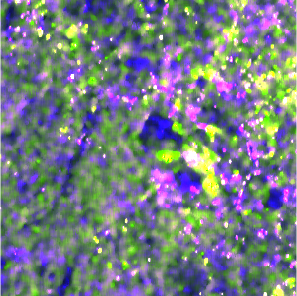}&
		\includegraphics[width=0.15\textwidth]{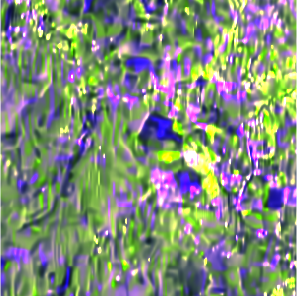}&
		\includegraphics[width=0.15\textwidth]{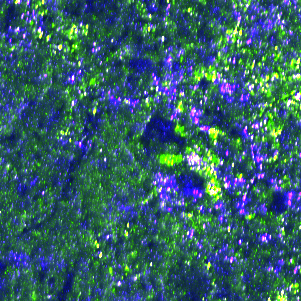}&
		\includegraphics[width=0.15\textwidth]{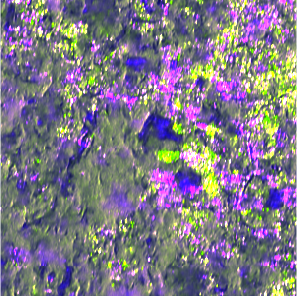}&
		\includegraphics[width=0.15\textwidth]{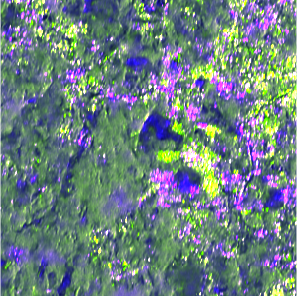}\\

	{(a)}&{(b)}& {(c)} &{(d)}& {(e)} & {(f)}  
		
	\end{tabular}

	\caption{Covariance matrix estimation result for a baseline method and CV-deSpeckNet for a 200 × 200 and 300 × 300 subset for test image 1 indicated on (Figure~\ref{fig:Fig5}). We show the color composite (red: $C_{11}$, green: $C_{22}$, blue: $C_{11}/C_{22}$) for (a) $\hat{C}$ (input), (b) Extended Lee sigma, (c) MuLoG (d) RV-FCN , (e) deSpeckNet (f) CV-deSpeckNet. } \label{fig:Fig3}
\end{figure*}

\section{Experimental Setup}
 
\subsection{Training}
\noindent Since the complex arithmetic used in the building blocks of complex-valued neural network can be simulated with real-valued arithmetic \cite{trabelsi1705deep}, the input data to the network  was prepared by vectorizing the Sentinel-1 $2 \times 2$ covariance matrix into a 6 channel image representing the real  and imaginary parts of the upper triangular elements of the Hermitian positive semi-definite covariance matrix. The diagonal elements of the coavariance matrix are real-valued intensity images so for mathematical convenience we added $0i$ to represent the imaginary part for these images. We also used a temporal average of 18 covariance matrices synthesized from 18 images of the same region to prepare  ${C}$ used as a proxy for the reference covariance matrix.  We assumed polarimetric stationarity that the dominant scattering mechanism of the scene is not changing within the considered time span. Since, 18 images are too few to synthesize a relatively noise free covariance matrix, we perform additional spatial filtering using the MuLoG framework proposed in \cite{deledalle2017mulog}.\\

 \noindent The networks were trained using Adam optimization method. Both $FCN_{cov}$ and $FCN_{noise}$ consists of 17 complex-valued convolutions with 48 filters. Complex-valued Batch normalization was used for every complex-valued convolution layer except the first and prediction layer. The networks were trained for 50 epochs with a learning rate of $10^{-3}$ for 30 eopchs and an additional 20 epochs with a rate of $10^{-4}$. We set the weight ($\mu$) of the $L_{cov}$ to 100 and the weight ($\xi$) of $L_{Noise}$ was given a value of 1. To apply this, we used a training set of 58,368 randomly selected patches of size $40 \times 40 \times 6$ representing the real and imaginary part of the unique upper triangular elements of the covariance matrix. A mini-batches of 64 samples and a weight decay factor of $5\times 10^{-4}$ was used.  We trained the network using the Keras complex library \cite{dramsch2019complex} on Google Colab platform using a Tesla T4 GPU.  \\
 
\noindent To assess the performance of CV-deSpeckNet, we compared it with extended Lee sigma filter \cite{lee2014polarimetric}, MuLoG \cite{deledalle2017mulog}, real-valued FCN (RV-FCN) that used real-valued building blocks and deSpeckNet \cite{mullissa2020despecknet}. RV-FCN was trained using the same architecture as $\mathrm{FCN}_{cov}$. { CV-deSpeckNet and deSpeckNet were trained using same number of parameters. In both cases unsupervised fine tuning \cite{mullissa2020despecknet} was not applied.} {Extended Lee Sigma and MuLoG are both unsupervised methods, which require only 1 second and 2.36 minutes, respectively. The training of RV-FCN required 1.3 hours, whereas deSpeckNet and CV-deSpeckNet required 2.6 hours and 7.01 hours, respectively.} 
 
\begin{figure*}
	\centering
	
		\begin{tabular}{cccccc}
		\includegraphics[width=0.15\textwidth]{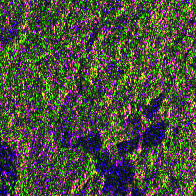}&
		\includegraphics[width=0.15\textwidth]{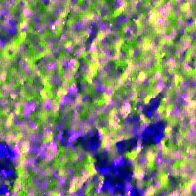}&
		\includegraphics[width=0.15\textwidth]{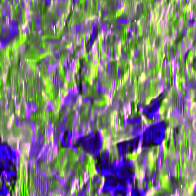}&
		\includegraphics[width=0.15\textwidth]{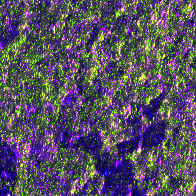}&
		\includegraphics[width=0.15\textwidth]{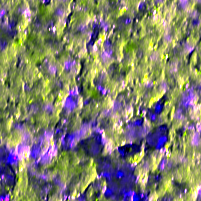}&
		\includegraphics[width=0.15\textwidth]{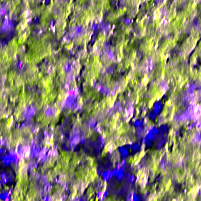}\\

	{(a)}&{(b)}& {(c)} &{(d)}& {(e)} & {(f)}  
		
	\end{tabular}

	\caption{{Covariance matrix estimation result for baseline methods and CV-deSpeckNet for  a $200 \times 200$ subset of test image 2. We show the color composite (red: $C_{11}$, green: $C_{22}$, blue: $C_{11}/C_{22}$) for (a) $\hat{C}$ (input), (b) Extended Lee sigma, (c) MuLoG (d) RV-FCN , (e) deSpeckNet (f) CV-deSpeckNet. }}\label{fig:Fig4}
\end{figure*}

\subsection{Quality metrics}

\noindent  To evaluate the performance of CV-deSpeckNet and the state of the art PolSAR despeckling methods quantitatively, we used temporally  averaged  images  obtained  in  the same  areas as the test images.  {The quality metrics used to evaluate the performance of the compared methods in the presence of full reference label image} are the peak signal to noise ratio (PSNR), structural similarity index (SSIM) \cite{mullissa2020despecknet}, despeckling gain (DG) \cite{di2013benchmarking}. These metrics are averaged for the diagonal elements of the covariance matrix which are real-valued intensity images. Furthermore, to evaluate the performance of CV-deSpeckNet and compared methods we use the absolute error of the polarimetric scattering mechanism ($\tilde{\alpha}$), entropy ($\tilde{H}$) and anisotropy ($\tilde{A}$) \cite{lee2004unsupervised} \cite{mullissa2018polarimetry} derived from the full covariance matrix.  \\

  \begin{table}[tbh]
	\begin{center}
	\scriptsize
		\setlength\tabcolsep{2pt}
		\begin{tabular}{lcccccccc} \hline \hline
Area	& Method  & PSNR & SSIM & DG & ENL & $\tilde{\alpha}$ & $\tilde{H}$ & $\tilde{A}$  \\ \hline
\multirow{4}{*}{\rotatebox{90}{~~Test image 1~~}}	&Lee sigma   & 26.43  & 0.76 &0.11 & 54.61 & 3.47 & 0.062 & 0.058 \\
                &MuLog   & 26.93  & 0.79 &0.54 & 46.05 & 2.48 & 0.054 & 0.051\\
                &RV-FCN   & 20.13  & 0.48 &-6.14 & 14.49& 8.22 & 0.08 & 0.07\\
                &deSpeckNet   & 26.82  & 0.78 & 1.49 & 46.99 & 1.75 & 0.041 & 0.038\\ 
	            &CV-deSpeckNet   & 27.06  & 0.79 & 1.05 & 81.46 & 1.75 & 0.04 & 0.03\\ 
	            &CV-deSpeckNet (test)  & 31.12  & 0.87 & 5.31 & 177.58 & 1.34 & 0.031 & 0.029\\\hline
\multirow{4}{*}{\rotatebox{90}{~~Test image 2~~}}	&Lee sigma  & 30.93  & 0.73 & 5.09&17.45 & 3.70 & 0.06 & 0.05\\
	            &MuLog  & 32.58  & 0.78 &9.66 & 12.15 & 3.12 & 0.065 & 0.063\\
	            &RV-FCN & 21.03  & 0.33 & -1.04&10.06  & 8.56 & 0.14 & 0.11\\
	            &deSpeckNet & 30.92  & 0.72 & 6.72 & 7.40  & 1.28 & 0.03 & 0.028\\
	            &CV-deSpeckNet & 34.49  & 0.82 & 12.73 & 29.08  & 1.25 & 0.027 & 0.025 \\ 
	            &CV-deSpeckNet (test) & 37.28  & 0.88 & 16.72 & 86.81 & 0.9 & 0.021 & 0.020 \\\hline \hline
		\end{tabular}\\
	\end{center}
	\caption{PSNR, SSIM, DG, ENL values averaged for the two diagonal elements of the covariance matrix and the $\tilde{\alpha}$, $\tilde{H}$ and $\tilde{A}$ values derived from the full covariance matrix. CV-deSpeckNet (test)  refers to CV-deSpeckNet fine-tuned on a temporally averaged image for the test scene used as an upper bound. }\label{tbl:Tab5}
\end{table} 

\noindent To evaluate the despeckling ability of CV-deSpeckNet and the compared methods without a reference data,  we used visual inspection as a qualitative measure and the equivalent number of looks (ENL) derived in a homogeneous region as a quantitative measure for comparison.

\section{Results and Discussion} 
\subsection{Qualitative Assessment}

\noindent In test image 1, both extended Lee sigma and MuLog resulted in a smooth output but on close inspection many artefacts can be observed. Whereas,{ RV-FCN failed to remove most of the speckle from homogenous regions and  preserve the subtle features in the image}. The filtered output still maintained the noisy appearance of the single-look covariance matrix. This was due to the insufficient number of training samples, as progressive increase of the training set improved its performance.  CV-deSpeckNet and deSpeckNet performed better than the other methods in removing speckle while preserving subtle features such as edges and point scatterers (Figure~\ref{fig:Fig3}). Since, test image 1 is acquired over the training area with a different realization of speckle, the results indicates that the multi-stream architecture is more robust in learning underlying speckle noise.  \\

\noindent In test image 2, both extended Lee sigma filter and MuLog still maintained a smooth output whereas {RV-FCN still failed remove speckle and preserve subtle features. This was expected as the image scene and noise distribution was expected to vary from the data it was trained on.}  However, CV-deSpeckNet resulted in the best output as less noise was observed in the estimated image composite and most subtle features were preserved  (Figure~\ref{fig:Fig4}). {This clearly show the generalization capability of multi-stream complex-valued architectures.} In the test regions, for all methods, supervised tuning of the model was not performed. The reference temporally averaged covariance matrix was only used to derive the quantitative quality metrics. 

\subsection{Quantitative Assessment}

\noindent The potential of CV-deSpeckNet is further exemplified by the improvement of the quantitative metrics defined in IV.B.  In test image 1,  CV-deSpeckNet achieved the highest PSNR, SSIM and ENL values compared to extended Lee sigma, MuLog, RV-FCN and deSpeckNet (Table~\ref{tbl:Tab5}). It also achieved the lowest absolute error for ($\tilde{\alpha}$), entropy ($\tilde{H}$) and anisotropy ($\tilde{A}$) values. CV-deSpeckNet achieved the highest result in all quality metrics  when comparing the quantitative results in test image 2.  This indicates the robust feature representation learning ability within multi-stream complex-valued FCN's. To establish the upper bound for the test image 1 and 2, we applied supervised tuning of CV-deSpeckNet in the test images using the temporally averaged reference data. These results show that CV-deSpeckNet, ever without using supervised tuning on the test image, is able to reach comparable performance to a network tuned on a reference test image (Table~\ref{tbl:Tab5}).  

\section{Conclusion}

\noindent {We have presented CV-deSpeckNet, a multi-stream complex-valued  fully convolutional network architecture, that is able to learn a model suitable for despeckling and effective estimation of dual-polarized covariance matrix. Our experiments on test images, obtained over a different regions, confirm the robustness of CV-deSpeckNet.} CV-deSpeckNet proved to be effective in estimating the covariance matrix while preserving the image quality with a modest sized training data. It was also able to learn robust feature representation that was able to adapt to a new test image. It provided better estimation results than the state of the art methods and its real-valued counterpart {and a comparable performance obtained by a FCN models optimized with temporally averaged images}. 

\section*{Acknowledgement}

\noindent This work was partly funded through the Global Forest Watch-World Resources Institute (GFW-WRI) Radar for Detecting Deforestation (RADD) project and the U.S. government SilvaCarbon Program. This paper contains modified Copernicus Sentinel data (2015-2020).

\bibliographystyle{unsrt}
\bibliography{References.bib}

\end{document}